\documentclass[sigconf]{acmart}

\AtBeginDocument{%
  }

\setcopyright{acmlicensed}
\copyrightyear{2025}
\acmYear{2025}
\setcopyright{cc}
\setcctype{by-nc}
\acmConference[ICAIF '25]{6th ACM International Conference on AI in Finance}{November 15--18, 2025}{Singapore, Singapore}
\acmBooktitle{6th ACM International Conference on AI in Finance (ICAIF '25), November 15--18, 2025, Singapore, Singapore}\acmDOI{10.1145/3768292.3770354}
\acmISBN{979-8-4007-2220-2/2025/11}

\usepackage{amsmath}  
\usepackage{graphicx}
\usepackage{hyperref}
\usepackage[utf8]{inputenc}
\usepackage[T1]{fontenc}
\usepackage{microtype}
\usepackage{multirow}
\usepackage{url}
\usepackage{CJK}

\begin{document}

\title{Query Generation Pipeline with Enhanced Answerability Assessment for Financial Information Retrieval}

\author{Hyunkyu Kim}
\orcid{0009-0005-7858-8136}
\authornote{Contributed equally to this research.}
\affiliation{%
  \institution{Kakaobank}
  \city{Seongnam-si}
  \country{Republic of Korea}}
\email{conor.k@lab.kakaobank.com}

\author{Yeeun Yoo}
\orcid{0000-0001-7370-3178}
\authornotemark[1]
\affiliation{%
  \institution{Kakaobank}
  \city{Seongnam-si}
  \country{Republic of Korea}}
\email{anny.ye@lab.kakaobank.com}

\author{Youngjun Kwak}
\orcid{0009-0004-4805-8786}
\authornote{Corresponding author.}
\affiliation{%
  \institution{Kakaobank}
  \city{Seongnam-si}
  \country{Republic of Korea}}
\email{vivaan.yjkwak@lab.kakaobank.com}


\begin{abstract}
As financial applications of large language models (LLMs) gain attention, accurate Information Retrieval (IR) remains crucial for reliable AI services. 
However, existing benchmarks fail to capture the complex and domain-specific information needs of real-world banking scenarios. Building domain-specific IR benchmarks is costly and constrained by legal restrictions on using real customer data. To address these challenges, we propose a systematic methodology for constructing domain-specific IR benchmarks through LLM-based query generation. As a concrete implementation of this methodology, our pipeline combines single and multi-document query generation with an enhanced and reasoning-augmented answerability assessment method, achieving stronger alignment with human judgments than prior approaches. Using this methodology, we construct KoBankIR, comprising 815 queries derived from 204 official banking documents. Our experiments show that existing retrieval models struggle with the complex multi-document queries in KoBankIR, demonstrating the value of our systematic approach for domain-specific benchmark construction and underscoring the need for improved retrieval techniques in financial domains.

\end{abstract}

\ccsdesc[500]{Information systems~Query reformulation}
\ccsdesc[500]{Information systems~Relevance assessment}
\ccsdesc[500]{Information systems~Test collections}

\keywords{Query Generation, Automatic Query Evaluation, Financial Information Retrieval, Banking Information Retrieval, Multi-document Retrieval, Benchmark Dataset}


\maketitle
\section{Introduction}
As AI-driven financial services continue to evolve, the demand for accurate and domain-specific Information Retrieval (IR) systems has significantly increased \cite{FinIR, choi2023conversational, wang2023csprdfinancialpolicyretrieval}. Financial institutions, particularly banks, require sophisticated IR capabilities to handle diverse customer inquiries ranging from product information to policy clarifications. These systems are crucial in delivering reliable and context-aware responses to banking customers, ultimately enhancing service quality. However, existing financial IR benchmarks suffer from several limitations, making them unsuitable for real-world banking applications \cite{choi2023conversational, wang2023csprdfinancialpolicyretrieval, kim2024fcmrrobustevaluationfinancial, hwang2025twiceadvantageslowresourcedomainspecific}. 

First, current financial IR research has largely focused on retrieving information from structured financial reports, news articles, and stock market data for market analysis and decision-making \cite{choi2023conversational, wang2023csprdfinancialpolicyretrieval}. In contrast, customer-facing financial services require IR systems capable of handling various institutional documents. Second, existing datasets lack multi-hop and multi-document queries in an IR setting, limiting their applicability to complex real-world financial information needs \cite{kim2024fcmrrobustevaluationfinancial}. Third, the linguistic and cultural specificity of financial services remains a challenge. Korean financial institutions, for instance, face unique challenges in developing IR systems that can handle domain-specific terminology and regulatory language \cite{hwang2025twiceadvantageslowresourcedomainspecific}. The availability of domain-specific IR benchmarks and supporting models is significantly limited, hindering the development and evaluation of effective retrieval systems.

To address these limitations, we introduce a query generation pipeline that systematically produces multi-document queries for financial IR system, coupled with an enhanced answerability assessment method. Our methodology enables the creation of realistic multi-document queries that better reflect actual customer information needs in banking environments. To demonstrate the effectiveness of our approach, we apply this pipeline to produce KoBankIR, a benchmark specifically designed for Korean-language banking information retrieval. KoBankIR comprises 815 queries derived from 204 official bank product disclosures, representing real-world banking interactions. Unlike existing financial IR datasets, KoBankIR explicitly incorporates multi-document queries generated through our proposed pipeline, mirroring how customers actually seek and retrieve information across multiple banking documents.

\begin{figure*}[ht]
\centering
  \includegraphics[width=0.8\textwidth]{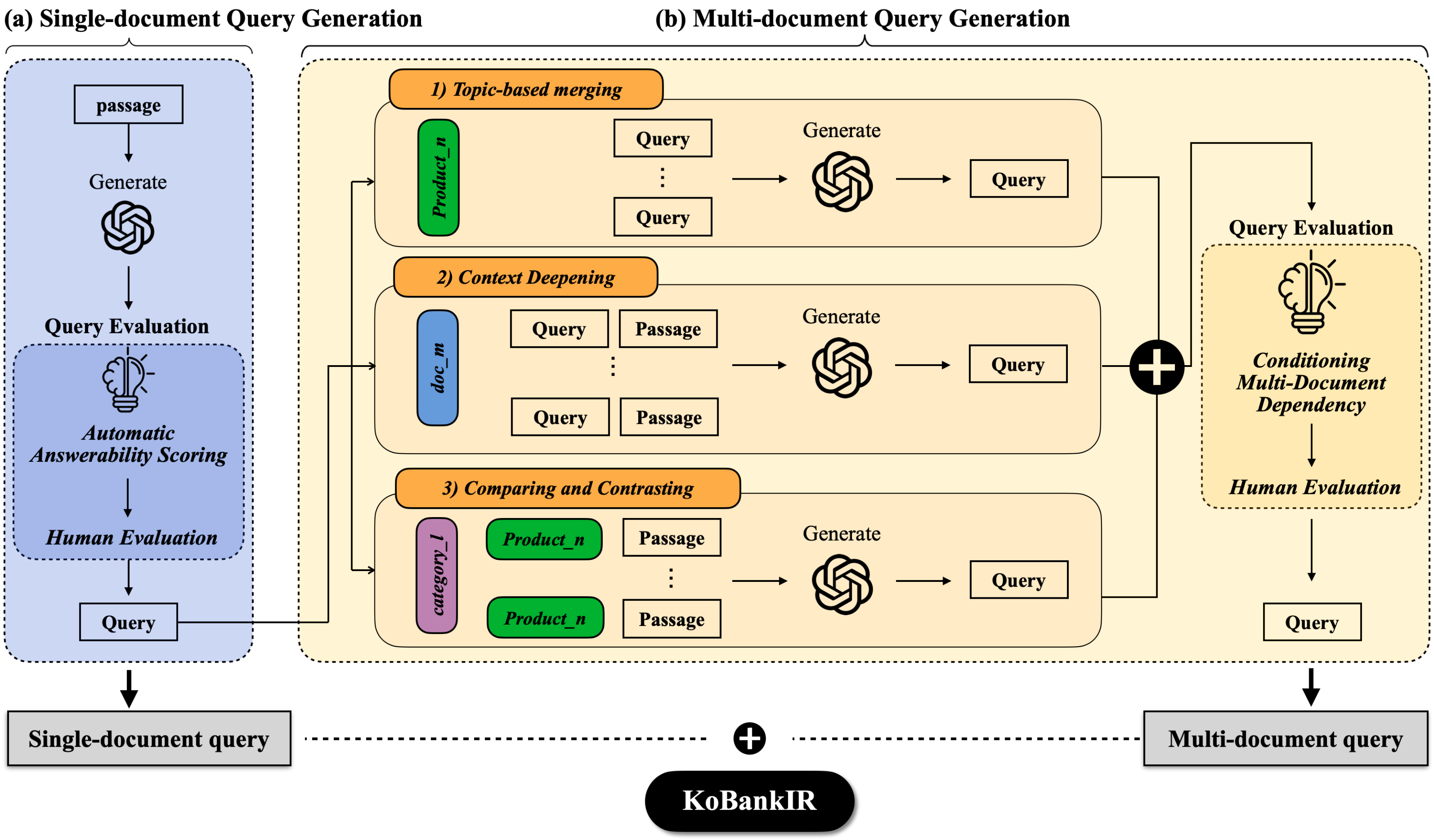}
  \caption{The query generation pipeline. (a)~Single-document query generation. (b)~Multi-document query generation, comprising (1)~Topic-based merging, (2)~Context deepening, and (3)~Comparing and contrasting. The generated queries undergo evaluation to ensure quality and answerability.}
  \label{fig:querygenerpipe}
\end{figure*}

As shown in Figure~\ref{fig:querygenerpipe}, our pipeline comprises three primary steps: (i) single-document query generation 
using domain-specific prompts, (ii) multi-document query generation through topic-based merging, context deepening, and comparing \& contrasting, and (iii) automated evaluation of query answerability using a improved reasoning-augmented method. The query answerability assessment method provides a more robust approach by offering better alignment with human evaluation than previous automatic scoring methods, such as G-EVAL \cite{liu2023gevalnlgevaluationusing}, which was analyzed in the QGEval study \cite{fu2024qgevalbenchmarkingmultidimensionalevaluation}. This is achieved through a think-augmented evaluation technique based on a reasoning-trained model (DeepSeek-R1-Distill-Qwen-14B model \cite{deepseekai2025deepseekr1incentivizingreasoningcapability}). This ensures both the quality of the dataset and its applicability to real-world banking scenarios.

We conduct comprehensive experiments to evaluate our methodology using multilingual retrieval models on the KoBankIR dataset, which we developed by applying our query generation pipeline to 204 official banking product disclosures, resulting in 815 high-quality queries. Our experiments demonstrate that existing retrieval models struggle to handle the domain-specific and complex multi-document queries generated by our pipeline. This reveals both the challenging nature of real-world financial IR tasks and the limitations of current IR models, while validating the effectiveness of our query generation approach.

Our key contributions are as follows:

\begin{itemize}
    \item We develop a structured query generation pipeline that integrates single-document and multi-document query generation with automated answerability evaluation for financial IR systems.
    \item We introduce a reasoning-augmented answerability assessment method that significantly improves upon existing automatic evaluation methods by providing better alignment with human judgment.
    \item We demonstrate the effectiveness of our approach through comprehensive experiments on Korean banking data, establishing new benchmarks for financial IR system evaluation and revealing significant challenges for existing retrieval models.
\end{itemize}

\section{Related Works}

\subsection{Information Retrieval Datasets}
IR datasets are typically structured as query-document pairs, enabling retrieval models to be evaluated across various domains. While general-domain IR datasets have contributed significantly to retrieval advancements \cite{thakur2021beir, geva2021did, zhang2023miracl, bandarkar2023belebele, valentini2024messirvelargescalespanishinformation}, domain-specific datasets are essential for handling specialized tasks, such as legal \cite{ma2021lecard, lotfi2024bilingual, li2024lecardv2, chen2025slard}, scientific \cite{wang2023dorismaescientificdocumentretrieval, ding2024pdfmvqadatasetmultimodalinformation}, and financial information retrieval \cite{FiQA}.


In the financial domain, existing datasets primarily emphasize retrieving information and Question Answering(QA). CSPRD \cite{wang2023csprdfinancialpolicyretrieval} supports policy retrieval, whereas REFinD \cite{kaur2023refind} and FinRED \cite{sharma2023finreddatasetrelationextraction} focus on relation extraction rather than document retrieval. Similarly, FinQA \cite{chen2021finqa}, FinTextQA \cite{chen2024fintextqadatasetlongformfinancial}, and DocFinQA \cite{reddy2024docfinqalongcontextfinancialreasoning} prioritize answer generation over retrieval performance.

Despite these advancements, existing financial IR datasets remain heavily focused on structured reports, stock market data, and numerical reasoning, and fail to address real-world banking scenarios in which customers retrieve and integrate information across multiple documents \cite{choi2023conversational, wang2023csprdfinancialpolicyretrieval}. We propose a systematic methodology for constructing domain-specific IR benchmarks through query generation and enhanced quality assessment. 


 
\subsection{Multi-document, Multi-hop Questions}


Several multi-hop, multi-document QA datasets exist, such as HotpotQA \cite{yang2018hotpotqa}, ComplexWebQuestions \cite{talmor2018webknowledgebaseansweringcomplex}, 2WikiMultiHopQA \cite{ho2020constructing}, and FanOutQA \cite{zhu2024fanoutqa}. These evaluate reasoning by requiring multiple inference steps over linked documents. However, although these datasets support complex multi-hop, multi-document and multi-modal reasoning, they are primarily designed for structured QA tasks rather than for document retrieval. Moreover, most of these benchmarks are built on general-domain corpora like Wikipedia, making them less suitable for domain-specific tasks such as financial information retrieval.

\subsection{Automatic Query Generation Evaluation}
Ensuring the quality of generated queries is a crucial step in constructing a reliable information retrieval benchmark. Although human evaluation is the most accurate method for assessing the query quality, it is time-consuming and unscalable for filtering large candidate sets. Therefore, an effective automatic evaluation metric is essential for prescreening generated queries efficiently and maintaining high data quality \cite{liu2023gevalnlgevaluationusing}. 

However, evaluating automatically generated questions remains a significant challenge in this field. QGEval \cite{fu2024qgevalbenchmarkingmultidimensionalevaluation} conducted a comprehensive analysis, which reveals a notable misalignment between automatic evaluation metrics and human judgment. Their findings indicate that even G-EVAL, the most human-aligned metric among existing methods, achieved only a modest Pearson correlation coefficient of 0.36. This highlights the considerable gap between current automatic evaluation approaches and the human assessments of generated questions. To address this misalignment, we introduce a reasoning-augmented evaluator that demonstrates significantly stronger alignment with human evaluations. 

\section{Methodology}

\begin{figure}[t!]
\centering
  \includegraphics[width=0.7\linewidth]{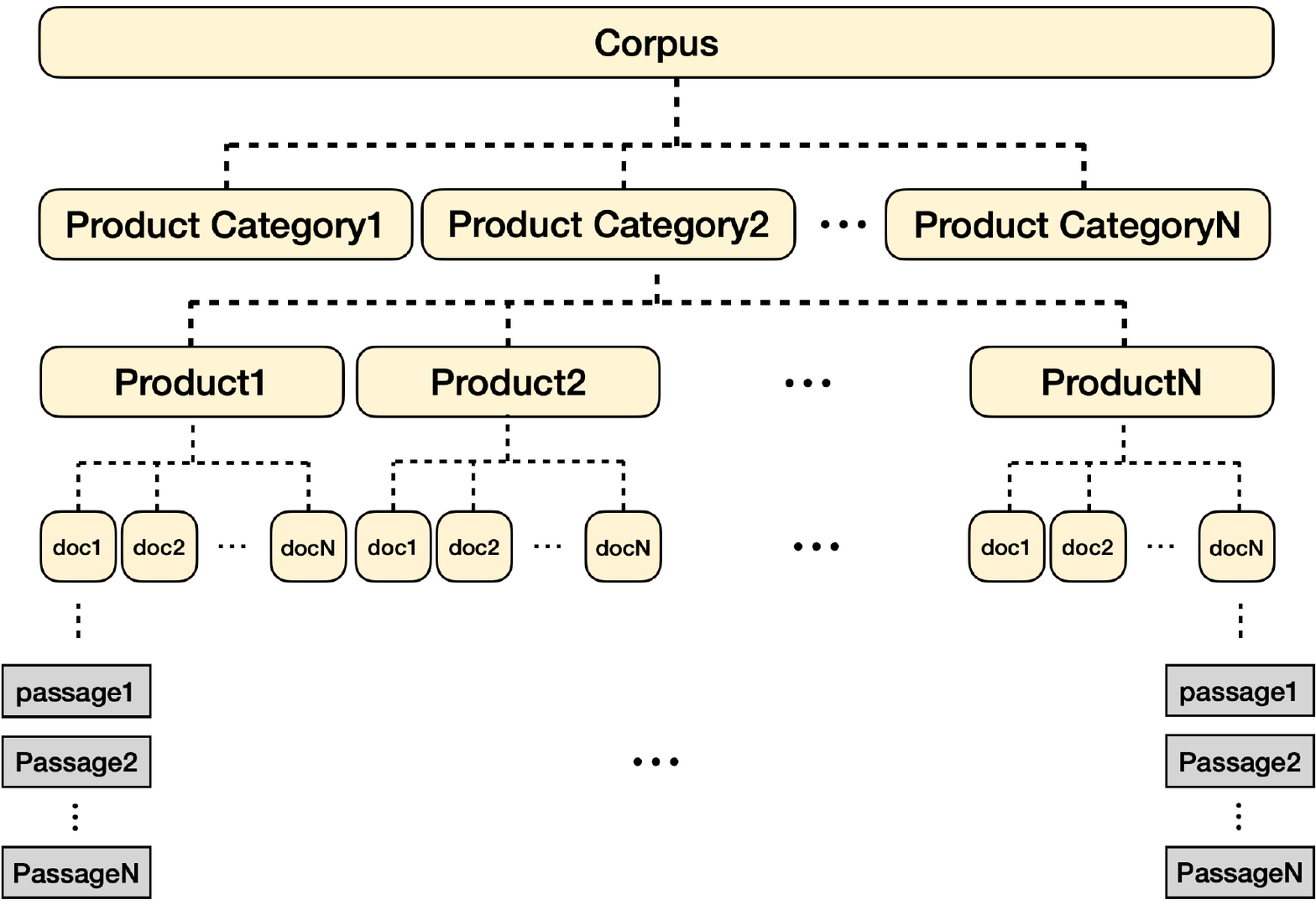}
  \caption{Hierarchical Structure of the Financial Document Dataset.}
  \label{fig:hierarchical_structure}
\end{figure}

In this section, we present the query generation pipeline for constructing KoBankIR. We outline data collection and preprocessing, describe our query generation pipeline, and introduce query evaluation method.


\subsection{Data Collection \& Preprocessing}

The KoBankIR dataset is constructed from 204 real-world financial documents provided by a Korean bank, covering 48 financial products such as time deposits, installment savings, and other banking services. These documents include product disclosures, terms and conditions, and policy documents. To support query generation, the documents were segmented into 1,794 coherent passages through a chunking process that reflects the hierarchical structure of the financial documents, as illustrated in Figure~\ref{fig:hierarchical_structure}.

\subsection{Query Generation Pipeline}
\label{sec:section3_2}
Owing to legal and ethical constraints prohibiting the use of actual customer inquiries, we implement a query generation pipeline to ensure compliance while maintaining relevance to real-world financial applications. As illustrated in Figure~\ref{fig:querygenerpipe}, the pipeline comprises two key components: (a) single-document query generation and (b) multi-document query generation.

\subsubsection{Single-document Query Generation}
Using the constructed passages, we generate a query for each single-document passage. We leverage GPT-4o as a query generator with a structured prompt template. Formally, let $G$ represents the generator, and the initially generated query set $Q= \{{q_1,q_2, \dots , q_n}\}$ be defined as: 
\begin{equation}
\begin{split}
    q_i=G(T_{single}(p_i)),
  \label{eq:Generator}
\end{split}
\end{equation}
where $T_{single}$ denotes the prompt template designed for single-document query generation, and $p_i$ represents an individual passage within the passage set $P=\{{p_1,p_2,\dots,p_{|P|}}\}$. Each query $q_i$ is initially generated without filtering, implying that some queries may not be fully answerable based on their respective passages. Additionally, we add metadata to the queries, including the product name under the label "the product held by the user" to link each query to a specific product.

To ensure that only high-quality queries are included in the final dataset, we apply an query evaluation using reasoning-augmented answerability assessment method that evaluates whether a query can be answered based on its corresponding passage. Only the queries deemed answerable through this evaluation are retained in the single-document query set. Further details can be found in Section \ref{sec:section3_3}. 

\subsubsection{Multi-document Query Generation}
After generating and filtering single-document queries based on query evaluation, we systematically construct multi-document queries that better reflect real-world banking IR scenarios in which users frequently seek information distributed across multiple sources. Through an intensive review of actual customers` inquiries, we define three types of multi-document query generation approaches: 1) Topic-based merging, 2) Context deepening, and 3) Comparing and contrasting. These three approaches are illustrated in the right panel of Figure~\ref{fig:querygenerpipe}, and examples for each are shown in Table \ref{tab:exampleofmultiquery}.

\begin{CJK}{UTF8}{mj}
\begin{table}[t!]
\centering
\begin{center}
\scriptsize
\renewcommand{\arraystretch}{1.2}
\begin{tabular}{p{0.2\columnwidth} p{0.7\columnwidth}}
\toprule
\textbf{Type} & Example \\
\midrule
\textbf{Topic-based merging} & (English) Where can I find the basic terms and conditions of Bank Special Sale Term Deposit, and what is the effective date of this special offer?\\ & 
(Korean) 이 은행의 특별판매 정기예금의 예금거래기본약관을 어디서 볼 수 있나요? 그리고 이 특약의 시행일은 언제인가요? \\
\midrule
\textbf{Contextual deepening} & (English) Do you check if a new home is purchased during the loan and how many homes a spouse owns at the time of loan origination? \\ & (Korean) 대출 기간 동안 신규 주택을 구입하는지 확인하나요? 그리고 대출 실행 시 배우자가 보유한 주택 수를 확인하나요? \\
\midrule
\textbf{Comparing~and contrasting} & (English) How are the prepayment penalties different for this Bank's Personal Loan and Household Loan products? \\ & (Korean) 이 은행의 개인 대출과 주택담보대출 상품의 중도상환 수수료는 어떻게 다르나요?\\
\bottomrule
\end{tabular}
\caption{Examples of multi-document query}
\label{tab:exampleofmultiquery}
\end{center}
\end{table}
\end{CJK}

\paragraph{1) Topic-based merging}
As shown in the top section of Figure~\ref{fig:querygenerpipe}(b), this method merges two or three single-document queries are merged into a single sentence. Single-document queries are randomly sampled from those associated with the same financial products. Appropriate conjunctions (e.g., "and," "also," "in addition," and "however") are incorporated to maintain natural linguistic integration.

To construct $k$-document queries, we first group single-document queries according to their associated financial product. For each $product$, we sample $k$ queries from the  corresponding query set, as follows:
\begin{equation}
    \{q_1, \cdots, q_k\} = \delta(Q^{{product}_i}, k),
\end{equation}
where $\delta$ is random sampling function and, $Q^{{product}_i}$ represents the set of single-document queries belonging ${product}_i$.

Finally, by using a structured prompt template $T_{merge}$, the query generator $G$ reformulates them into a multi-document query:
\begin{equation}
    q_{merged}=G(T_{merge}(\{q_1, \cdots, q_k\}),
\end{equation}

This allows us to construct queries that reflect real-world scenarios, in which users naturally ask multiple questions together.

\paragraph{2) Context deepening}
In this merging approach, illustrated in the middle section of Figure~\ref{fig:querygenerpipe}(b), we sample multiple query-passage pairs from the single-document queries, and the generator, $G$, considers both queries and passages during the merging process. This generates outputs, which simulate realistic financial consultations.

First, we identify the queries that originate from the same document. For each $document$, we sample $k$ of query-passage pairs:
\begin{equation}
    \{(q_1, p_1), \cdots, (q_k, p_k)\} = \delta(Q^{{document}_i}, k), 
\end{equation}
where $Q^{{document}_i}$ represents the set of single-document queries and their associated passages within ${document}_i$.

Using a structured prompt template $T_{deep}$ for deepening queries, the query generator, $G$, reformulates these query-passage pairs \newline $\{(q_1, p_1), \cdots, (q_k, p_k)\}$ into a deepened multi-document query:
\begin{equation}
    q_{deep}=G(T_{deep}(\{(q_1, p_1), \cdots, (q_k, p_k)\})),
\end{equation}
As shown in Table~\ref{tab:exampleofmultiquery}, these questions require multiple reasoning steps to answer.

\paragraph{3) Comparing and contrasting}
As shown in the bottom section of Figure~\ref{fig:querygenerpipe}(b), this method constructs multi-document queries by identifying comparable passages across different products within a financial product category. Unlike the previous methods, where queries are initially generated from individual passages and then merged or deepened, this approach directly utilizes the passages as generator inputs. The objective is to create queries that emphasize the similarities and differences between key financial concepts and policies.

First, we choose one category and then select several products within it. From each chosen product, we sample one passage. For example, selecting three products yields a set of three passages.

The generator, $G$, creates a query by comparing and contrasting the selected passages:

\begin{align}
    p_1 &= \delta(P^{{category}_i}, 1) \nonumber\\ 
    &\vdots \\
    p_k &= \delta(P^{{category}_i}, 1) \nonumber
\end{align}

\begin{equation}
    q_{comp}=G(T_{comp}(\{p_1, \cdots, p_k\})),
\end{equation}
where $T_{comp}$ is a structured prompt template designed to encourage contrastive and comparative reasoning and, $P^{{category}_i}$ represents the set of passages within ${category}_i$. These queries reflect real-world scenarios in which users compare financial options or policies across products.



\subsection{Query Evaluation}
\label{sec:section3_3}

\begin{figure}[t!]
\centering
  \includegraphics[width=0.95\linewidth]{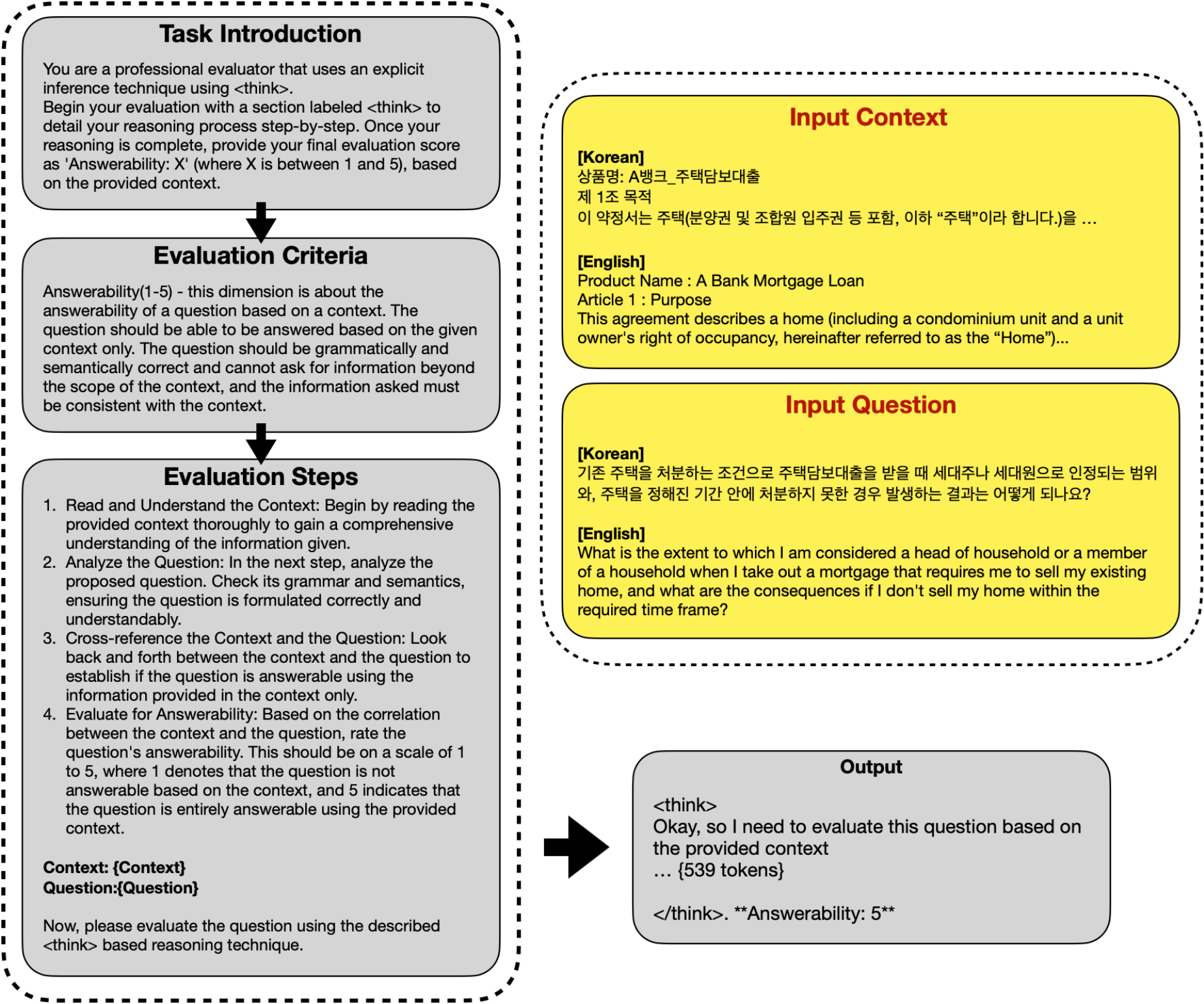}
  \caption{The process of reasoning-augmented evaluation method.}
  \label{fig:ThinkEval process}
\end{figure}

\subsubsection{Automatic Answerability Scoring}

To evaluate query answerability, we introduce a reasoning-augmented evaluator built on \textit{DeepSeek-R1-Distill-Qwen (7B and 14B)} which guides the model through explicit "Think" steps. Following the findings of the QGEval paper, which reported that G-EVAL-4 achieved the best performance among its competitors yet still exhibited limited alignment with human judgments (Pearson's $\rho$ = 0.36 on the QGEval dataset; see Table~\ref{tab:QGEval_results}), we replicated part of their experimental setup and extended it with additional experiments tailored to our approach. By incorporating explicit reasoning steps into the evaluation, we achieved substantially higher alignment with human judgments on the QGEval benchmark (Pearson's $\rho$ = 0.60 vs 0.36; Table~\ref{tab:QGEval_results}).

\begin{table}
    \centering
    \begin{tabular}{lrr}
    \toprule
    \textbf{Metric} & \multicolumn{1}{c}{\textbf{$\rho $}} & \multicolumn{1}{c}{\textbf{$\tau$}} \\
    \midrule
    G-Eval-gpt4     & 0.356                                & 0.169                                    \\
    G-Eval-gpt4o    & 0.302                                & 0.183                                    \\
    G-Eval-gpto1          & 0.542                                & 0.295                   \\
    Ours(7B)    & 0.299                                & 0.187                                    \\
    \textbf{Ours(14B)}   & \textbf{0.598}                                & \textbf{0.304}                                    \\
    \bottomrule
    \end{tabular}
\caption{Pearson ($\rho$) and Kendall-Tau ($\tau$) correlations between automatic evaluation metrics and human assessments for query answerability. G-EVAL-gpt4 results are from the original QGEval study, while G-EVAL-gpt4o, G-EVAL-gpt-o1, and our proposed reasoning-augmented evaluation methods using DeepSeek models (7B and 14B) were evaluated in this work. Our method with DeepSeek-14B achieves the highest correlation with human judgments.}
\label{tab:QGEval_results}
\end{table}

\begin{table}[t!]
    \centering
    \begin{tabular}{lrr}
    \toprule
    \textbf{Metric} & \multicolumn{1}{c}{\textbf{$\rho $}} & \multicolumn{1}{c}{\textbf{$\tau$}} \\
    \midrule
    G-Eval-gpt4     & 0.647                                & 0.451                                    \\
    \textbf{Ours(14B)}   & \textbf{0.775}                                & \textbf{0.565}                                    \\
        \bottomrule
    \end{tabular}
\caption{Pearson ($\rho$) and Kendall-Tau ($\tau$) correlations with human judgments on the KoBankIR dataset for G-EVAL-gpt4 and our proposed reasoning-augmented evaluation approach using DeepSeek-14B. Our method achieves substantially higher alignment with human evaluations.}
\label{tab:GEval_Thinkeval_results_kobankir}
\end{table}

Surprisingly, with just a minor prompt modification, the proposed evaluator outperformed all baselines in matching human judgments-even though it is much smaller than GPT-4o \cite{openai2024gpt4technicalreport}. Figure~\ref{fig:ThinkEval process} illustrates the complete "Think" reasoning chain, showing how each query is evaluated step by step. We also tested GPT-o1 under identical conditions. Because OpenAI's API limits extended reasoning paths, we reverted to the simpler G-EVAL prompts for GPT-o1, which consequently showed lower alignment with human judgments compared to our reasoning-augmented approach (see Table~\ref{tab:QGEval_results}).

Although we use Pearson and Kendall-tau correlations to measure alignment with human judgments-as done in QGEval-our primary focus is on its performance for query filtering. Only 6 of 450 queries (1.33\%) receive high answerability scores ($\ge$ 4) despite low human ratings ($<$ 2). This result shows that our assessment method rarely overestimates query quality compared to human judgments, demonstrating its reliability as an automated filtering mechanism with a minimal false-positive rate.

Since QGEval was not originally designed for Korean, so its direct adaption to our Korean banking dataset may raise concerns. To address this, we conducted a complementary human evaluation following the QGEval's answerability criteria on our data. From the initial pool of generated queries, we stratified samples into four bins-below 2, [2, 3), [3, 4), and 4 or above-and randomly selected 25 queries from each bin to form a set of 100 samples. These were then independently rated by the authors of this paper on a 1-3 answerability scale. The resulting correlations, shown in Table~\ref{tab:GEval_Thinkeval_results_kobankir}, confirm that our evaluator's higher alignment with human judgments holds true even in a Korean-language setting.

\subsubsection{Conditioning Multi-Document Dependency}
When synthesizing the multi-document queries, we observed that many of the generated queries could be answered using a single passage, contradicting the multi-document dependency condition. To address this issue, we apply an additional filtering mechanism.
We define a valid multi-document query as one in which the answerability score of the query with all supporting contexts combined is higher than the score when evaluated with any individual document alone. Formally, for a query with the corresponding passages, the following condition must hold:
\begin{equation}
  \label{eq:conditional_dependency}
  F(P_c,Q_c) > max(F(P_1,Q_c),\cdots, F(P_n,Q_c)),
\end{equation}
where $P_c$ represents the union of all relevant passages for the multi-document query $Q_c$. 
This condition requires that $Q_c$ be more answerable when provided with all passages rather than just a subset of them. By applying this criterion, we eliminated a substantial number of queries, ensuring the final dataset included only those that require information across multiple contexts.

\begin{figure}[t!]
\centering
  \includegraphics[width=1\linewidth]{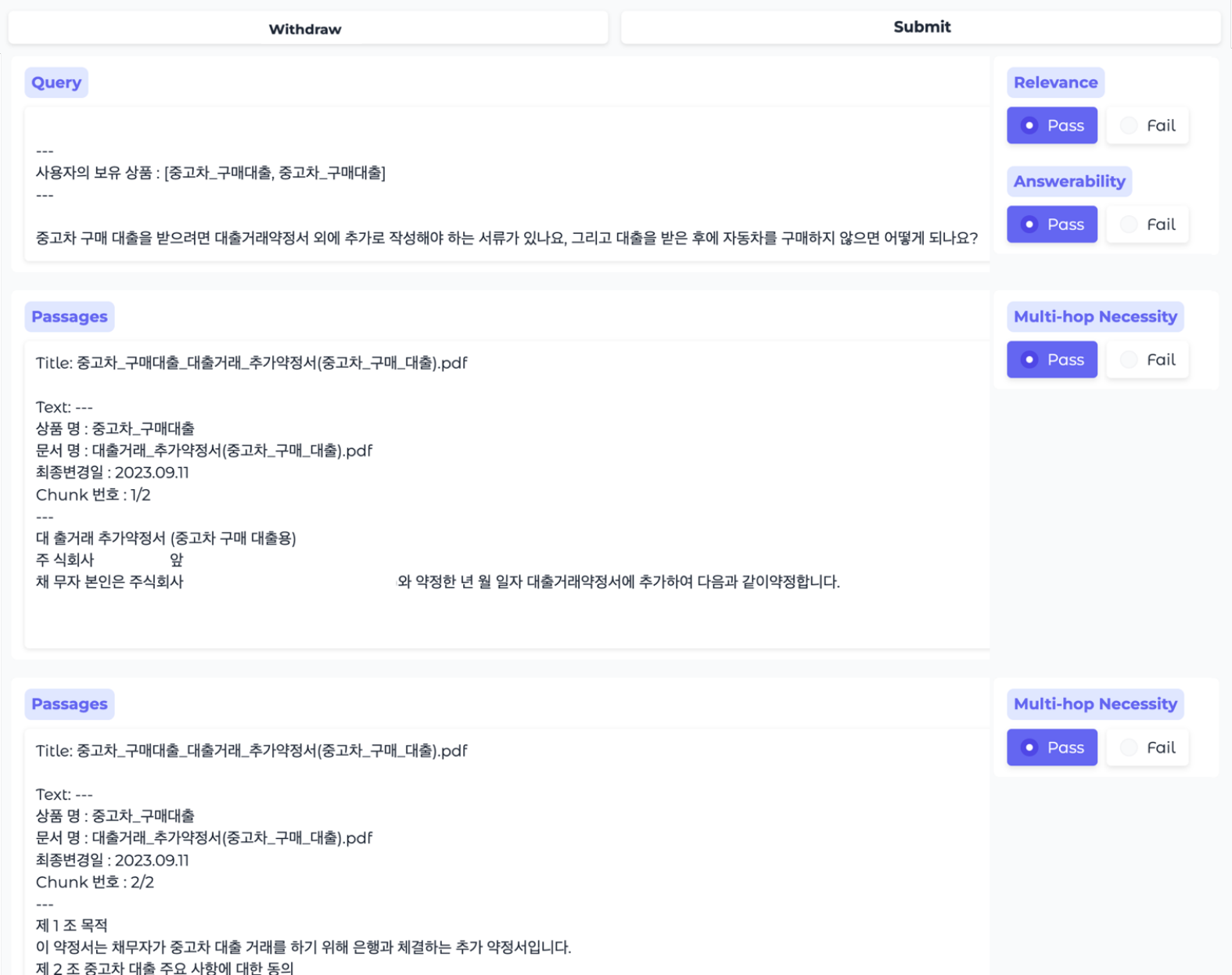}
  \caption{Human evaluation interface.}
  \label{fig:human_eval_interface}
\end{figure}

\subsubsection{Human Evaluation}
To ensure the quality of the dataset, we conducted a human evaluation on queries with an answerability score greater than 4.0, as determined by our reasoning-augmented assessment method. The purpose of this evaluation was to refine the dataset and ensure that only high-quality queries were included in the benchmark. The evaluation was conducted by five banking professionals, each with at least two years of financial industry experience, using a single review approach in which each query was evaluated by an expert reviewer to ensure efficient processing while maintaining quality standards.


\begin{itemize}
    \item \textbf{Relevance}: The query must be highly relevant to banking-related information retrieval tasks.
    \item \textbf{Answerability}: The query must be answerable based on the provided passage.
    \item \textbf{Multi-hop Necessity}: Queries should necessitate multi-document reasoning, and unrelated passages must be filtered out.
\end{itemize}

You may see the human evaluation tool in the Figure~\ref{fig:human_eval_interface}.

\begin{table}[h]
\small
\centering
\renewcommand{\arraystretch}{1.2}
\begin{tabular}{lrrrr}
\toprule
\textbf{Query Type} & \textbf{1-docs} & \textbf{2-docs} & \textbf{3-docs} & \textbf{4-docs} \\
\midrule
\textbf{Single-document Queries} & 457 & - & - & - \\
\midrule
\textbf{Multi-document Queries} & & & & \\
\quad Topic-based Merging       & - & 69  & 54  & 0   \\
\quad Context Deepening         & - & 59  & 35  & 2   \\
\quad Comparing and Contrasting & - & 93  & 31  & 15  \\
\bottomrule
\end{tabular}
\caption{Statistics of the KoBankIR dataset: the numbers indicate the number of queries for each type and number of supporting documents.}
\label{tab:data_stat2}
\end{table}

\begin{table*}[ht!]
\centering
\renewcommand{\arraystretch}{1.2}
\small
\begin{tabular}{lrrrrrrr}
\toprule
\multirow{2}{*}{}                                        & \multicolumn{1}{|l}{}        & \multicolumn{2}{|c|}{NDCG@}                      & \multicolumn{2}{|c|}{mAP@}                       & \multicolumn{2}{|c}{Recall@}                    \\ \hline
                                                         & \multicolumn{1}{|c|}{EmbSize} & \multicolumn{1}{|c}{5} & \multicolumn{1}{c|}{10} & \multicolumn{1}{|c}{5} & \multicolumn{1}{c|}{10} & \multicolumn{1}{|c}{5} & \multicolumn{1}{c}{10} \\ \hline\hline
\multicolumn{8}{c}{\textbf{Sparse retrieval models}}                                                                                                                                                        \\ \hline
BM25                                                     & \multicolumn{1}{|c|}{}        & 0.1768                & \multicolumn{1}{c|}{0.2063}                 & \multicolumn{1}{c}{0.1455}                & \multicolumn{1}{c|}{0.1579}                 & 0.2216                & 0.3038                 \\ \hline
BAAI/bge-m3 (\textit{Sparse})                    & \multicolumn{1}{|c|}{1024}                        & 0.5722                & \multicolumn{1}{c|}{0.6127}                 & 0.5047                & \multicolumn{1}{c|}{0.5263}                 & 0.6723                & 0.7783                  \\ \hline\hline
\multicolumn{8}{c}{\textbf{Dense retrieval models}}                                                                                                                                                                                             \\ \hline
Alibaba-NLP/gte-Qwen2-1.5B-instruct      & \multicolumn{1}{|c|}{1536}                        & 0.5650                & \multicolumn{1}{c|}{0.6105}                 & 0.5045                & \multicolumn{1}{c|}{0.5278}                 & 0.6540                & 0.7741                 \\ \hline
intfloat/multilingual-e5-large           & \multicolumn{1}{|c|}{1024}                        & 0.6165                & \multicolumn{1}{c|}{0.6545}                 & 0.5526                & \multicolumn{1}{c|}{0.5726}                 & 0.7069                 & 0.8061                  \\ \hline
intfloat/multilingual-e5-large-instruct & \multicolumn{1}{|c|}{1024}                        & 0.6180                & \multicolumn{1}{c|}{0.6625}                 & 0.5527                & \multicolumn{1}{c|}{0.5765}                 & 0.7141                & 0.8315                 \\ \hline
BAAI/bge-m3 (\textit{Dense})            & \multicolumn{1}{|c|}{1024}                        & {0.6452}       & \multicolumn{1}{c|}{{0.6848}}        & {0.5828}       & \multicolumn{1}{c|}{{0.6043}}        & {0.7342}       & {0.8387}        \\ \hline\hline
\multicolumn{8}{c}{\textbf{Multi-vector retrieval models}}                                                                                                                                                                                             \\ \hline
BAAI/bge-m3 (\textit{Multi})                   & \multicolumn{1}{|c|}{1024}                        & 0.6408                & \multicolumn{1}{c|}{0.6789}                 & 0.5796                & \multicolumn{1}{c|}{0.5997}                  & 0.7226                & 0.8228                 \\ \hline\hline
\multicolumn{8}{c}{\textbf{Hybrid approaches}}                                                                                                                                                                                                     \\ \hline
BAAI/bge-m3 (\textit{Sparse + Dense + Multi})                     & \multicolumn{1}{|c|}{1024}                        & 0.6565                & \multicolumn{1}{c|}{0.6917}                 & 0.5838                & \multicolumn{1}{c|}{0.6041}                 & 0.7267                & 0.8271                 \\ \hline
\textbf{BAAI/bge-m3 (\textit{Sparse + Dense})}                    & \multicolumn{1}{|c|}{\textbf{1024}}                        & \textbf{0.6950}        & \multicolumn{1}{c|}{\textbf{0.7109}}        & \textbf{0.6167}        & \multicolumn{1}{c|}{\textbf{0.6345}}        & \textbf{0.7723}       & \textbf{0.8528}        \\ \hline
BAAI/bge-m3 (\textit{Dense + Multi})                              & \multicolumn{1}{|c|}{1024}                        & 0.6377                & \multicolumn{1}{c|}{0.6749}                 & 0.5806                & \multicolumn{1}{c|}{0.6002}                 & 0.7246                & 0.8232       \\ \bottomrule         
\end{tabular}
\caption{Experimental results comparing sparse, dense, multi-vector, and hybrid retrieval models on NDCG, mAP, and Recall metrics. \textbf{Bold} denotes best score in each column. The results show that hybrid approach (BGE-M3 Sparse + Dense) consistently outperforms other models.}
\label{table:Experimental results}
\end{table*}

\subsection{Data Statistics}

As a result of human evaluation, we selected a total of 815 queries for the benchmark, including 358 multi-document queries and 457 single-document queries. Table~\ref{tab:data_stat2} shows the detailed statistics of single-document and  multi-document queries, which are categorized into three types: Topic-based Merging (123 queries), Context Deepening (96 queries), and Comparing and Contrasting (139 queries). These queries are distributed across different document combinations, ranging from 2-docs to 4-docs configurations. 

\section{Experiments}

\subsection{Baselines}
Sparse, dense, multi-vector retrieval models are prominent approaches for information retrieval \cite{chen2024bgem3embeddingmultilingualmultifunctionality}. In this section, we evaluate the three types of retrieval models using our benchmark to assess their performance in real-world banking scenarios. 


\paragraph{Sparse retrieval models}
Sparse methods like BM25 \cite{robertson2009probabilistic} are widely used due to their simplicity and interpretability \cite{ma2023enhancing, Mandikal2024Sparse}, but often fail to capture deeper semantic relationships \cite{Formal2023Towards}. Recent works have addressed this by enhancing term representations \cite{mallia2021learning}. We evaluate BM25 and BGE-M3 (\textit{Sparse}), a lexicon-based model with Korean support. Notably, BGE-M3 supports sparse, dense and multi-vector retrieval, making it a versatile choice for evaluating different retrieval paradigms.

\paragraph{Dense retrieval models}
Dense retrieval models encode documents and queries into dense vectors and match them based on similarity measures \cite{Luan2020Sparse, karpukhin-etal-2020-dense, xiong2020approximate, neelakantan2022textcodeembeddingscontrastive, Chen2023End-to-End, wang2024textembeddingsweaklysupervisedcontrastive, chen2024bgem3embeddingmultilingualmultifunctionality}. Dense retrieval models dominate current retrieval system research, particularly because the adoption of LLMs has significantly boosted their performance \cite{wang2023improving,lee2024nv}. Based on strong multilingual performance in MTEB \cite{muennighoff2023mtebmassivetextembedding}, we selected three representative models for our experiments: GTE-Qwen2-1.5B-instruct \cite{li2023towards}, Multilingual-e5-Large, Multilingual-e5-Large-instruct \cite{wang2024multilingual}, and BGE-M3.

\paragraph{Multi-vector retrieval models}
Multi-vector retrieval models enhance retrieval effectiveness by representing queries and documents with multiple vectors, thereby enabling more fine-grained semantic matching \cite{khattab2020colbertefficienteffectivepassage, dhulipala2024muveramultivectorretrievalfixed}. Although these approaches demonstrate improved performance compared to single-vector models through token-level interaction mechanisms, they encounter scalability challenges when applied to large-scale collections \cite{dhulipala2024muveramultivectorretrievalfixed}. In our experimental setup, we utilize BGE-M3 as a multi-vector retrieval model, primarily because of the limited availability of open-source models with official Korean language support.

We also evaluate hybrid retrieval strategies that combine sparse, dense, and multi-vector outputs to leverage the strengths of each representation.


\subsection{Experimental Settings}
Dense retrieval used cosine similarity for ranking, while sparse retrieval used lexical matching through sparse matrix multiplication. Multi-vector retrieval captured token-level similarity using MaxSim operations \cite{khattab2020colbertefficienteffectivepassage}. The hybrid approach used three weight configurations: (0.4, 0.3, 0.3), (0.7, 0.3, 0), and (0.5, 0, 0.5) for dense, sparse, and multi-vector retrieval respectively. The models were evaluated using NDCG \cite{jarvelin2002cumulated}, mAP, and Recall metrics.

\subsection{Experimental Results}

The complete experimental results on the baselines are presented in Table ~\ref{table:Experimental results}.

\textbf{Sparse retrieval models} (BM25 and BGE-M3 (\textit{Sparse})) exhibit limited effectiveness in banking scenarios. BM25 performs the worst among all baselines, with an NDCG@5 of 0.1768. In contrast, BGE-M3 (\textit{Sparse}) significantly outperforms BM25, achieving an NDCG@5 of 0.5722, demonstrating that lexicon-based sparse models benefit from deep learning enhancements. However, despite these improvements, sparse retrieval remains inferior to the dense and multi-vector approaches.

\textbf{Dense retrieval models} exhibit stronger ranking performance compared to the sparse and multi-vector retrieval models. Among the dense models, BGE-m3 (\textit{Dense}) achieves the best score with an NDCG@5 of 0.6452, surpassing both Alibaba-NLP/gte-Qwen2-1.5B-instruct (0.565) and intfloat/multilingual-e5-large (0.6165). 

\textbf{Multi-vector retrieval model} (BGE-M3 (\textit{Multi})) achieves an NDCG@5 of 0.6408, which is comparable to that of the best dense model.

Hybrid retrieval approaches that combine sparse, dense, and multi-vector representations yield the best overall performance. Notably, BGE-M3 (\textit{Sparse + Dense}) achieves an NDCG@5 of 0.6795, outperforming all other models, including the pure dense and multi-vector approaches. Additionally, it records the highest mAP@5 (0.6167) and Recall@10 (0.8663), indicating that the hybrid approaches simultaneously enhance ranking and recall. These results reinforce the idea that combining sparse and dense retrieval effectively balances lexical matching and semantic understanding, making it the most suitable approach for financial document retrieval.

However, even with the hybrid method achieving the best results, overall performance remains modest, highlighting significant room for improvement in financial IR techniques.

\paragraph{Performance by Document Count}
\label{sec:document_count_performance}

To further analyze performance differences by query type, we evaluate the best-performing model (BGE-M3 (\textit{Sparse + Dense})) on single- and multi-document queries. Table~\ref{tab:exp2} shows the NDCG@5 scores broken down by the number of supporting documents per query. While the model achieves strong performance on single-document queries, performance degrades as the number of documents increases, highlighting the greater challenge of multi-document retrieval.

\begin{table}[t!]
\centering
\begin{tabular}{lr}
\toprule
\textbf{Document count} & \textbf{NDCG@5} \\
\midrule
\textbf{Single-document Queries} & 0.749 \\
\midrule
\textbf{Multi-document Queries} & \\
\quad 2-documents   & 0.590   \\
\quad 3-documents   & 0.633   \\
\quad 4-documents   & 0.302   \\
\bottomrule
\end{tabular}
\caption{NDCG@5 performance of the best model (BGE-M3(\textit{Sparse + Dense})) by document count.} 
\label{tab:exp2}
\end{table}

\paragraph{Performance by Multi-document Query Type}
\label{sec:md_querytype_perf}

To better understand how retrieval performance varies across different types of multi-document queries, we evaluate the best-performing model (BGE-M3(\textit{Sparse + Dense})) on each query type. As shown in Table~\ref{tab:md_querytypes}, the model performs well on Topic-based Merging and Context Deepening queries, but shows a notable performance drop on Comparing and Contrasting queries. This suggests that comparative reasoning remains a challenging task in current retrieval models.

\begin{table}[t!]
\centering
\begin{tabular}{lr}
\toprule
\textbf{Query Type} & \textbf{NDCG@5}\\
\midrule
\textbf{Single-document Queries} & 0.749 \\
\midrule
\textbf{Multi-document Queries} & \\
\quad Topic-based Merging       & 0.709\\
\quad Context Deepening         & 0.696\\
\quad Comparing and Contrasting & 0.414\\
\bottomrule
\end{tabular}
\caption{NDCG@5 performance of the best model (BGE-M3(\textit{Sparse + Dense})) by query type. Performance on comparing and contrasting queries is notably lower.}
\label{tab:md_querytypes}
\end{table}

\section{Conclusion}

We propose a structured query generation pipeline that combines single-document and multi-document queries with automated answerability evaluation, enabling the construction of high quality information retrieval benchmarks. As part of this pipeline, we introduce a reasoning-augmented evaluator for query answerability, which improves the quality and reliability of the generated dataset.  By using this pipeline, we build KoBankIR, the first Korean-language benchmark for banking-domain information retrieval. Our experiments on KoBankIR highlight the limitations of current retrieval models in this domain, emphasizing the need for further research on more effective retrieval techniques.

\bibliographystyle{ACM-Reference-Format}
\bibliography{99_ref}






\end{document}